# Noise Error Pattern Generation Based on Successive Addition-Subtraction for Guessing Decoding

Ming Zhan, Member, IEEE, Zhibo Pang, Senior Member, IEEE, Kan Yu, Member, IEEE, Jing Xu, and Fang Wu

M. Zhan and J. Xu are with the College of Electronics and Information Engineering, Southwest University, Chongqing 400715, China (e-mail: zmdjs@swu.edu.cn, xujinggg@163.com).
Z. Pang is with the ABB Corporate Research, Forskargr" and 7, SE-721 78 Vasterås, Sweden, he is also a visiting professor with the KTH Royal Institute of Technology, 10044 Stockholm, Sweden (e-mail: pang.zhibo@se.abb.com, zhibo@kth.se).
K. Yu is with the Department of Computer Science and Information Technology, School of Engineering and Mathematical Sciences, La Trobe University, Bendigo Campus, Edwards Rd, Flora Hill VIC 3552, Australia (e-mail: kan.yu@hotmail.com).
F. Wu is with the School of Economics and Management, Southwest University, Chongqing 400715, China (e-mail: wufanghrbcf@163.com).

*Abstract*—Guessing random additive noise decoding (GRAND) algorithm has emerged as an excellent decoding strategy that can meet both the high reliability and low latency constraints. This paper proposes a successive addition-subtraction algorithm to generate noise error permutations. A noise error patterns generation scheme is presented by embedding the "1" and "0" bursts alternately. Then detailed procedures of the proposed algorithm are presented, and its correctness is also demonstrated through theoretical derivations. The aim of this work is to provide a preliminary paradigm and reference for future research on GRAND algorithm and hardware implementation.

## I. INTRODUCTION

To improve productivity, adapt to flexible manufacturing and reduce process costs, wireless technologies for control have become one of the fundamental research hotspots in future smart factories [1]. In some critical industrial control scenarios, such as machine-to-machine communications (M2M), power systems automation (PSA), and power electronics control (PEC), large amounts of data are rapidly exchanged between the central controller and the nodes in the form of short packets (refresh rate up to $10^5$ Hz), and stringent requirements are imposed on the Packets error rate (PER) and latency of short packet transmission [2]. Ultra-reliability and low latency communications (URLLC) for short packets is one of the research hotspots in the field of industrial wireless control [3].

In order to achieve high reliability and low latency, channel coding is considered to be a practical and feasible technical strategy [4]. A better transmission reliability can be obtained in the cost of a certain computational complexity and a controllable coding/decoding delay. Under short code transmission conditions, traditional linear block codes (e.g., BCH, RS codes) and convolutional codes CC, which have better error correction performance and lower decoding delay than Shannon limit codes (e.g., Turbo and LDPC codes), have been widely used in industrial wireless control [5]. Unfortunately, an Industrial production process is cyclical in nature, with frequent start/stop of electrical equipment, closure and disconnection of control contacts, or various arc discharges (of different intensities and durations) formed by processes such as welding [6]; and pulse interference introduced from the ground line fleeing into the RF signal of wireless control devices [7]. These highly correlated pulse disturbances are one of the main factors causing the failures of wireless transmissions, leading to misoperation of the control systems [8].

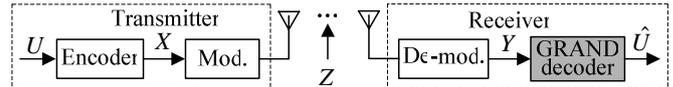

Fig. 1. Wireless communication system with GRAND decoder.

Guessing random additive noise decoding (GRAND) algorithm takes into account the memory characteristics of discrete channels, and regards the burst errors caused by the correlation between impulse interference and noise as an universal existence in discrete channels [9]. When random linear codes (RLCs) are applied, the decoding performance of the GRAND algorithm has a significant advantage of 2~3 dB over the classic structured coding schemes such as BCH and RM codes [10]. At present, there are very few researches on the GRAND algorithm [11], including the most critical noise error pattern generation methodology. Based on the given parameters such as the number of error bursts and the total length, this paper proposes an algorithm for successive add-subtract operation to generate noise error permutations, and theoretically proves the effectiveness of this method. Taking into account the correlation between the noise error bits in the discrete memory channel, by embedding the "1" and "0" bit bursts alternately, the generated noise error pattern can be directly decoded with the received sequence.

This paper is organized as follows. Section II introduces the basic idea of GRAND decoding algorithm, the noise error pattern, and its statistical probability distribution. Section III discusses the structure of noise error pattern generation and proposes a step-by-step addition and subtraction noise error pattern generation algorithm, and makes corresponding theoretical proofs based on detailed analysis, and proves its feasibility by examples. Conclusions of the study are given in Section IV.

## II. THE GUESSING RANDOM ADDITIVE NOISE DECODING

### A. Basic algorithm strategies

The GRAND algorithm directly points to the channel noise that causes the transmission error, and guesses the possible random noise sequence introduced by the channel [10] (referred to as the noise error pattern in this paper). The block diagram of a wireless communication system using GRAND decoder is shown in Fig. 1.

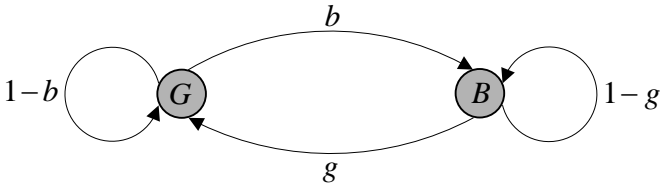

Fig. 2. State transition of a discrete channel with Markov chain.

Let the length of the encoded short packet be $N$, and the encoded sequence of the original packet be $X = x_1\, x_2 \cdots x_i \cdots x_N$, the wireless channel additive interference sequence is $Z = z_1\, z_2 \cdots z_i \cdots z_N$, the demodulation sequence is $Y = y_1\, y_2 \cdots y_i \cdots y_N$, the GRAND decoding output is $\hat{U}$, where $X$ is any sequence in the codeword space $C$ and the symbols $x_i$, $y_i$ and $z_i$ belong to the binary set $\{0,1\}$. The relationship between $X$, $Z$ and $Y$ is shown in Eq. 1.

$$Y = X \oplus Z \quad (1)$$

According to the statistical characteristics of discrete channels, the GRAND decoder generates binary noise error patterns $S = s_1\, s_2 \cdots s_i \cdots s_N$ in order of probability in descending order. Estimated sequence $\hat{X}$ of $X$ can be calculated by Eq. 2.

$$\hat{X} = Y \oplus S \quad (2)$$

For each calculated $\hat{X}$, compare and judge whether this sequence is a legal codeword in $C$, as shown in Eq. 3.

$$\begin{cases} \hat{X} \in C,\ output\ \hat{X}, decoding\ success \\ \hat{X} \notin C,\ generate\ new\ S, repeat\ Eq.\ (2) \end{cases} \quad (3)$$

If the decoding is successful, then $S=Z$ is considered correct. Since $\hat{X}$ is a legal code word in $C$, the estimated sequence $\hat{U}$ of $U$ is easily obtained according to the encoding rules and output as the decoding result. Otherwise, a new noise error pattern S is generated and Eq. (2) is repeated until the decoding is successful, or when the specified number of decoding times is exceeded, a decoding error is reported.

### B. Statistical probability of noise

The binary discrete channel can be described by a two-state Markov chain, as shown in Fig. 2 [12].

Suppose that the good state when there is no noise interference in the channel is $G$, the corresponding symbol in Z is "0", and the bad state when an error occurs in the state reversal is $B$, the corresponding symbol in Z is "1". The probability of state G transitioning to $B$ is $b$, the probability of state $B$ transitioning to $G$ is $g$, the probability of $G$ remaining in the original state is $1-b$, and the probability of $B$ remaining in the original state is $1-g$. The stationary bit-flip probability of the Markov chain is $b/(b+g)$, the correlation coefficient is $1-b-g$, and when $b+g=1$, it is a memoryless channel. Suppose the number of "1" symbol bursts in sequence Z is $m$ ($m \leq \lfloor N/2 \rfloor$), and the total number of all "1" symbols is $l_m$ ($l_m \geq m$), which is called $\{m, l_m\}$ in this paper a combination of noise parameters. For the specified $\{m, l_m\}$ parameter combination, according to the different distribution of the "1" burst in the sequence $Z$, it can be divided into the following three cases [10].

Case 1: The start and end bits in Z are both "0", the statistical probability $P_1(m, l_m)$ is calculated by Eq. (4).

$$P_1(m, l_m) = \frac{g}{1-b}\frac{(1-b)^N}{b+g}\left(\frac{bg}{(1-b)(1-g)}\right)^m \left(\frac{1-g}{1-b}\right)^{l_m} \quad (4)$$

Case 2: The start or end bit in Z is "1", the statistical probability $P_2(m, l_m)$ is calculated by Eq. (5).

$$P_2(m, l_m) = \frac{1-b}{g} P_1(m, l_m) \quad (5)$$

Case 3: The start and end bits in Z are both "1", the statistical probability $P_3(m, l_m)$ is calculated by Eq. (6).

$$P_3(m, l_m) = \left(\frac{1-b}{g}\right)^2 P_1(m, l_m) \quad (6)$$

Taking into account the actual situation of the channel, the stationary bit-flip probability $b/(b+g) < 1/2$, that is, $b < g$. When the noise is positively correlated, the correlation coefficient is $1-b-g > 0$, making $(1-b)/g > 1$, so the statistical probability relationship of the three cases is shown in Eq. (7).

$$P_1(m, l_m) < P_2(m, l_m) < P_3(m, l_m) \quad (7)$$

### III. NOISE ERROR PATTERN GENERATION ALGORITHM

#### A. Noise error pattern generation framework

As discussed in Section II, the noise error pattern $S$ is the key to the GRAND decoding algorithm, and $S$ is formed by alternately embedding "1" and "0" bursts. Corresponding to

The work was supported by the National Nature Science Foundation of China under Grant 61671390. (Corresponding author: Zhibo Pang.)

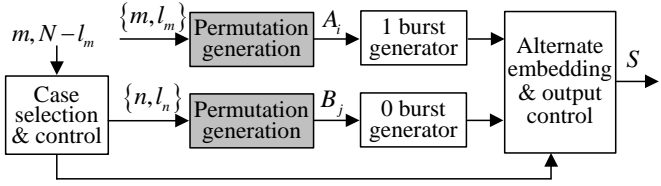

Fig. 3. Structure diagram of noise error pattern S generation.

cases 1, 2 and 3 in the "1" symbol parameter combination $\{m, l_m\}$, the "0" symbol parameter combination $\{n, l_n\}$ is $\{m+1, N-l_m\}$, $\{m, N-l_m\}$ and $\{m-1, N-l_m\}$ respectively. For $m$ "1" symbol bursts, define the number $a_{i,k}$ of "1" symbol in each burst to form a noise error permutation $A_i$ represented by Eq. (8), where $1 \leq k \leq m$ and satisfies $\sum_{k=1}^{m} a_{i,k} = l_m$.

$$A_i = \{a_{i,1}, a_{i,2}, \cdots, a_{i,k}, \cdots, a_{i,m}\} \quad (8)$$

There are many 1-noise error permutations that can meet the parameter combination $\{m, l_m\}$ requirement. Suppose the set of such permutations is $\mathbb{R}_1 = \{A_1, A_2, \cdots, A_i, \cdots, A_M\}$, and there are $M$ in total. Similarly, there are $W$ 0-noise error permutations $B_j = \{b_{j,1}, b_{j,2}, \cdots, b_{j,k}, \cdots, b_{j,n}\}$, $1 \leq j \leq W$ that satisfy the parameter combination $\{n, l_n\}$, constituting the set $\mathbb{R}_0 = \{B_1, B_2, \cdots, B_j, \cdots, B_W\}$. For each permutation $A_i$, generating all permutations $B_j$ corresponding to the 0 symbol parameter combination $\{n, l_n\}$. The elements $a_{i,k}$ and $b_{j,k}$ in $A_i$ and $B_j$ are respectively used to control the "1" burst and "0" burst generators to output the number $a_{i,k}$ of "1" and $b_{j,k}$ of "0", alternately embedded to form a noise error pattern $S$, the structure is shown in Fig. 3.

*B. The successive addition-subtraction algorithm*

$A_i$ and $B_j$ are the core of constructing the noise error pattern $S$. This paper takes $\mathbb{R}_1$ as an example to discuss the generation algorithm of arbitrary 1-noise error permutations $A_i$. The generation of $A_i$ can be attributed to the specific value of all elements $a_{i,k}$ in Eq. 8 for a specified parameter combination $\{m, l_m\}$. The successive addition-subtraction algorithm proposed in this paper includes the following steps.

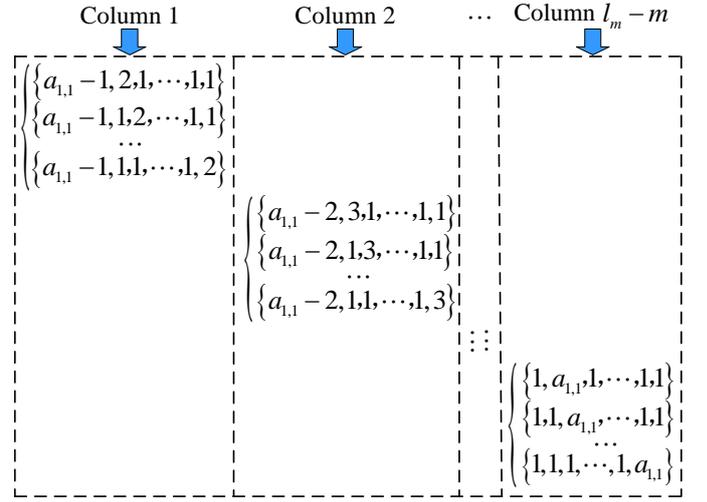

Fig. 4. Schematic diagram of successive add-subtract operation.

Step 1: Initialization of the first noise error permutation $A_1 = \{l_m - (m-1), 1, 1, \cdots, 1\}$.

Step 2: For $A_1$, the first element $a_{1,1}$ is successively subtracted from $h = 1, 2, \cdots, a_{1,1} - 1$, and the subtracted value is successively added to the subsequent elements to form a new permutation. This successive add-subtract operation process is shown in Fig. 4. For the convenience of discussion, the $(l_m - m) \times (l_m - m)$ cells included in the dashed box are combined as a $H$ grid. A cell is formed by the permutation of the first element that is equal. There are total $l_m - m$ cells generated in this step, which are located on the diagonal of the $H$ grid, and each cell includes $m-1$ permutations.

Step 3: Without loss of generality, for the $k^{th}$ permutation $A_i$ in the $h^{th}$ row and $h^{th}$ column cell of the $H$ grid in Fig. 4 ($1 \leq h \leq l_m - m - 1$, $1 \leq k \leq m-2$), the first element $a_{i,1} = a_{1,1} - h > 1$, and the values of $m-(k+1)$ elements after the element $a_{i,k+1} = 1+h$ are all 1. Its general form is represented by Eq. (9) ($1 \leq m-(k+1) \leq m-2$).

$$A_i = \left\{ a_{1,1} - h, \overbrace{1, \cdots, 1}^{k-1}, 1+h, \overbrace{1, 1, \cdots, 1, 1}^{m-(k+1)} \right\} \quad (9)$$

For the permutation $A_i$ that satisfies the general form of Eq. 9 in the $H$ grid of Fig. 4, the first element $a_{i,1} = a_{1,1} - h$ is successively subtracted by 1, 2, ... , $(a_{1,1} - h) - 1$ and then the subtracted value is successively added to the $m-(k+1)$ elements at the end to form a new permutation. According to the value of the first element of the new permutations, they are sequentially arranged in the cells corresponding to the $h^{th}$ row and $(h+1)^{th}$, $(h+2)^{th}$, ... , $(h+(a_{1,1}-h-1))^{th}$ columns of the $H$ grid. Considering that $a_{1,1} = l_m - (m-1)$, the columns $h+1$, $h+2$, ... , $h+(a_{1,1}-h-1)$ are the columns $h+1$, $h+2$, ... , $l_m - m$, and the value of the first element of all permutations in each column is also equal. The successive add-subtract operation process of the general form is shown in Fig. 5.

The subtracted value 1, 2, ... , $(a_{1,1}-h)-1$, are successively added to the element whose value is 1 after $a_{i,k+1} = 1+h$ in Eq. 9. The reason is as follows: For permutation $A'$, let the number to be added and subtracted is $1 \leq c \leq (a_{1,1}-h)-1$, if added to the position of $a_{i,f} = 1$ before the element $a_{i,k+1} = 1+h$, a new permutation is obtained, as shown in Eq. 10.

$$A' = \{a_{1,1}-h, 1, \cdots, a_{i,f}=1, 1, \cdots, 1, a_{i,k+1}=1+h, 1, \cdots, 1\}$$
$$\rightarrow add-subtract\ c \qquad (10)$$
$$= \{a_{1,1}-h-c, 1, \cdots, a_{i,f}=1+c, 1, \cdots, 1, a_{i,k+1}=1+h, 1, \cdots, 1\}$$

Then, there must exist another permutation $A'' = \{a_{1,1}-c, 1, \cdots, a_{i,f}=1+c, 1, \cdots, 1, a_{i,k+1}=1, 1, \cdots, 1\}$. By adding and subtracting $h$ to $A''$ in step 3, a new permutation shown in Eq. 11 will be obtained.

$$A'' = \{a_{1,1}-c, 1, \cdots, a_{i,f}=1+c, 1, \cdots, 1, a_{i,k+1}=1, 1, \cdots, 1\}$$
$$\rightarrow add-subtract\ to\ h \qquad (11)$$
$$= \{a_{1,1}-c-h, 1, \cdots, a_{i,f}=1+c, 1, \cdots, 1, a_{i,k+1}=1+h, 1, \cdots, 1\}$$

The results of Eq. 10 and Eq. 11 are the same. In order to ensure that new permutations are not generated repeatedly, this paper selects to perform the successive add-subtract operation after element $a_{i,k+1}$, and writes Eq. 9 into a more general form of Eq. 12.

$$A_i = \left\{ a_{1,1} > 1, \overbrace{a_{i,2}, \cdots, a_{i,k+1}}^{k}, \overbrace{1, 1, \cdots, 1, 1}^{m-(k+1)} \right\} \qquad (12)$$

Step 4: For the new permutations generated in step 3, step 3 is repeated if they conform to the general form of Eq. 12, until all newly generated permutations do not conform to the general form. When the step-by-step addition and subtraction operation ends, all permutations $\{A_1, A_2, \cdots, A_i, \cdots, A_M\}$ of the parameter combination $\{m, l_m\}$ are generated.

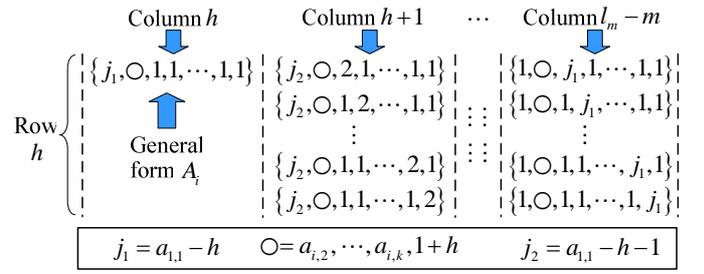

Fig. 5. Schematic diagram of successive add-subtract operation for general form.

Proof: The successive add-subtract algorithm of steps 1 to 4 give all permutations subject to the $m$ and $l_m$ parameters.

1) When the value of the first element is $a_{1,1} = l_m - (m-1)$, the sum of the remaining $m-1$ elements is $l_m - (l_m - (m-1)) = m-1$, such that each element can only have a value of 1. Such a permutation has only one $A_1$.

2) When the value of the first element of the permutations is $l_m - (m-1) - 1 = a_{1,1} - 1$, the sum of the remaining $m-1$ elements is $m$, and only one element has a value of 2. There are $m-1$ possibilities, corresponding to the $m-1$ permutations generated by adding 1 one by one from the $2^{nd}$ to the $m^{th}$ element, and located in the $1^{st}$ row and $1^{st}$ column of the $H$ grid in Fig. 4.

3) Without loss of generality, when the value of the first element of the permutation is $a_{1,1} - h$ ($2 \leq h \leq l_m - m$), the operand $h$ is successive added and subtracted to form new permutations in the following cases.

Define $\overrightarrow{\{e_1, e_2\}}$ to denote the permutation satisfying Eq. 12 in the $e_1^{th}$ row and $e_1^{th}$ column cell of the $H$ grid, performs the successive add-subtract operation to $e_2$ one by one, and the generated new permutations are located in the $e_1^{th}$ row and $(e_1 + e_2)^{th}$ column cell of the $H$ grid, as shown in Eq. (13).

$$\overrightarrow{\{e_1, e_2\}} \triangleq \begin{cases} A_i\ in\ e_1^{th} \times e_1^{th}\ cell\ of\ H, add-subtract\ e_2 \\ new\ permutations\ in\ e_1^{th} \times (e_1 + e_2)^{th} \end{cases} \qquad (13)$$

Similarly, a more general operation for multivariates $\overrightarrow{\{e_1, e_2, \cdots, e_{k-1}, e_k\}}$ means for permutation satisfying Eq. 12 in the $e_1^{th}$ row and $\left(\sum_1^{k-1} e_i\right)^{th}$ column cell of $H$, performs successive add-subtract to $e_k$, while the generated new permutations are located in the $e_1^{th}$ row and $\left(\sum_1^k e_i\right)^{th}$ column cell of the $H$ grid.

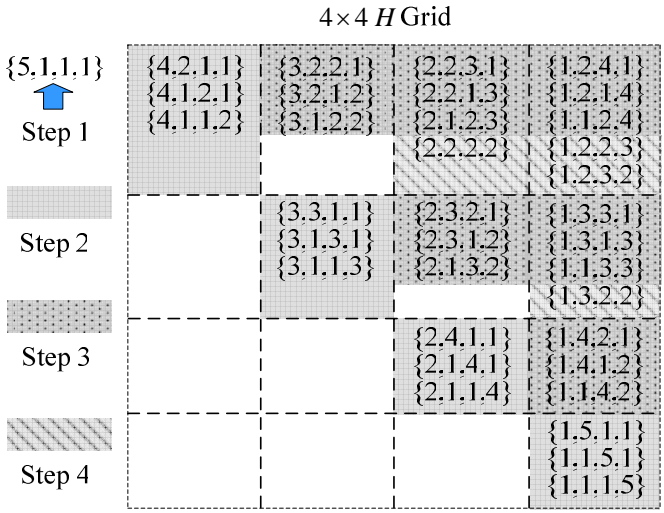

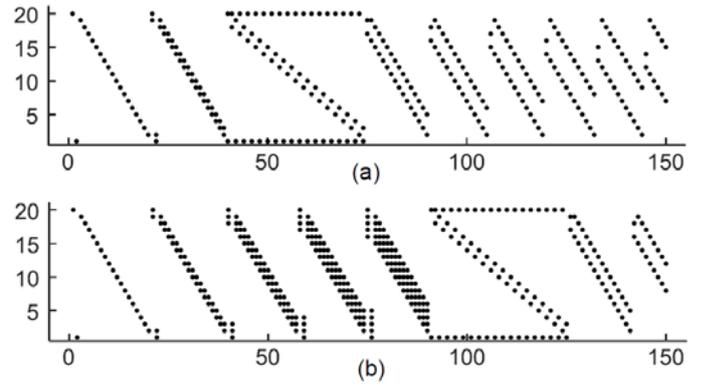

Fig. 7. Examples of generated noise error patterns: (a) $\Delta l = 0$, and (b) $\Delta l = 3$.

Fig. 6. Example of noise parameter combination $\{4,8\}$ permutations generation.

3-1) $h$ is decomposed into $(h)$, and $h$ is added one by one to the $m-1$ elements with value 1 at the end of $A_1$, corresponding to the $m-1$ permutations in the $h^{th}$ row and $h^{th}$ column cell of the $H$ grid.

3-2) The $h$ decomposition $(h-1, 1)$ corresponds to the $\overrightarrow{\{h-1,1\}}$ operation, and the newly generated permutations are located in the $(h-1)^{th}$ row and $h^{th}$ column cell of the $H$ grid.

3-3) The $h$ decomposition $(h-2, 2)$ and $(h-2, 1, 1)$ corresponds to the $\overrightarrow{\{h-2,2\}}$ and $\overrightarrow{\{h-2,1,1\}}$ operation, respectively. Then the new permutations are located in the $(h-2)^{th}$ row and $h^{th}$ column cell of the $H$ grid.

And so on, until $h$ decomposes into:

$(h-(h-1), h-1)$, corresponding to $\overrightarrow{\{1, h-1\}}$;

$(h-(h-1), 1, h-2)$, corresponding to $\overrightarrow{\{1,1,h-2\}}$;

$(h-(h-1), 2, h-3)$, $(h-(h-1), 1, 1, h-3)$, corresponding to $\overrightarrow{\{1,2,h-3\}}$, $\overrightarrow{\{1,1,1,h-3\}}$;

…

$(h-(h-1), h-2, 1)$, $(h-(h-1), 1, h-3, 1)$,

$(h-(h-1), h-3, 1, 1)$, …, $\left(h-(h-1), \overbrace{1, \cdots, 1}^{h-1}\right)$,

corresponding to $\overrightarrow{\{1, h-2, 1\}}$, $\overrightarrow{\{1,1, h-3, 1\}}$,

$\overrightarrow{\{1, h-3, 1, 1\}}$, …, $\overrightarrow{\{1, \overbrace{1, \cdots, 1}^{h-1}\}}$.

After all of the above successive add-subtract operations are completed, the first element value of the newly generated permutations is $a_{1,1} - h$, and they are all located in the $1^{st}$ row and $h^{th}$ column cell of the $H$ grid.

From the above proof, the successive addition-subtraction algorithm generates all permutations of the parameter combination $\{m, l_m\}$ starting from the initialized first permutation $A_1$, and arranges them in the $l_m - m$ columns of the $H$ grid in the order of the first element value from highest to lowest.

Example 1, calculating the permutations of the noise parameter combination $\{m, l_m\} = \{4, 8\}$ according to the successive addition-subtraction algorithm.

The initialization of the first line $A_1 = \{5,1,1,1\}$, step 1~4 execution process is shown in Fig. 6.

The noise error pattern $S$ generation structure in Fig. 3 is based on successive addition-subtraction algorithm to generate permutations $A_i$ and $B_j$. The parameter $\Delta l$, which reflects the channel memory characteristics, are calculated according to [10]. Subsequently, this parameter can be used to order the parameter combinations $\{m, l_m\}$. Let the code length $N = 20$, the first 150 noise error patterns $S$ for $\Delta l = 0$ and $\Delta l = 3$ are given in Fig. 7. The horizontal coordinates indicate the statistical probabilities in decreasing order and the vertical coordinates indicate the corresponding pattern $S = s_1\ s_2\ \cdots\ s_i\ \cdots\ s_{20}$, where "•" corresponds to the "1" symbol and blank corresponds to the "0" symbol.

## IV. CONCLUSIONS

The GRAND algorithm can effectively improve the reliability of short packet communication under impulsive interference channels and help reduce the decoding delay. Based on the analysis of the alternating embedded noise error pattern generation structure, this paper proposes a successive addition-subtraction algorithm to construct all permutations satisfying "1" symbol parameter combination $\{m, l_m\}$ and "0" symbol parameter combination $\{n, l_n\}$, and confirms the effectiveness of the algorithm through a rigorous mathematical proof. As the most critical noise error pattern generation algorithm in GRAND decoding, this paper provides a practical strategy for the hardware implementation of GRAND decoder.

ACKNOWLEDGMENT

The work was supported by the National Nature Science Foundation of China under Grant 61671390. (Corresponding author: Zhibo Pang.)